\documentclass[preprintnumbers,amsmath,amssymb]{revtex4-2}
\usepackage{graphicx}

\begin{document}

\title{Relevant spontaneous magnetization relations for the triangular and the cubic lattice Ising model}
\author{Tuncer Kaya }
\email{ tkaya@yildiz.edu.tr}
\affiliation{ Department of Physics, Yildiz Technical University, 34220 Davutpa\c sa, Istanbul, Turkey}

\begin{abstract}
The spontaneous magnetization relations for the 2D triangular and
the 3D cubic lattices of the Ising model are derived by a new
tractable easily calculable mathematical method. The 
result obtained for the triangular lattice is compared with the already
available result to test and investigate the relevance the new
mathematical method. From this comparison, it is seen that the
agreement of our result is almost the same or almost equivalent to
the previously obtained exact result. The new approach is, then,
applied to the long-standing 3D cubic lattice, and the
corresponding  expression for the spontaneous magnetism is
derived. The  relation obtained is compared with the already
existing numerical results for the 3D lattice. The essence of the
method going to used in this paper is based on exploiting the main
characteristic of the order parameter of  a second order phase
transition which provides a more direct physical insight into the
calculation of the spontaneous magnetization of the Ising model.

\end{abstract}

\maketitle
\section{\label{sec:level1}Introduction}

The Ising model is an approximate mathematical  model system to
describe and  investigate the  physically very cumbersome problem
of phase transition phenomena.  It was originated with Lenz
\cite{Lenz} in 1920 and was subsequently investigated by his
student Ising \cite{Ising}. Over a long period of time, it is seen
that the model unifies the study phase transitions in systems as
diverse as ferromagnet, gas-liquid, binary alloys, and so on. It
may be the simplest mathematical model to analyze and investigate
phase transitions theoretically. Its application to realistic
systems such as the 3D lattices is, however, still too cumbersome
to work with the conventional mathematical approaches used for the
2D lattices. Therefore, there is no exact result obtained from
this mode for the 3D lattices. In other words, the 3D model has
withstood challenges and remains, to this date, an outstanding
unsolved problem.  Even the application of the usual method to  2D lattices is very
formidable and very daunting. I think, therefore, that a new
mathematically more tractable and easily calculable method for the
treatment of these lattices may be interesting. Before starting to
introduce our mathematical approach to calculate the order
parameter or the spontaneous magnetization of the triangular and
the cubic lattices, it might be appropriate to introduce first
some important steps in the mathematical treatment of these
problems.

The existence of the critical point, meaning the existence of
second order phase transition,  was first proven by Kramers and
Wannier \cite{Kramers} using the method of a dual transformation.
The qualitative calculation of Kramers and Wannier was verified by
Onsager \cite{16} and \cite{Kaufman}. The existence of
criticality led to investigate and calculate the average
magnetization. Although, the spontaneous magnetization of the
Ising model on rectangular lattice was first calculated by Onsager
\cite{Beale}, he never bothered to publish his derivation
\cite{Keh}. Yang was the first person who published his derivation
of the average magnetization of the square lattice Ising model
\cite{Yang1952}. His result is
$<\sigma>=[1-\sinh(2K)^{-4}]^{\frac{1}{8}}$. Since the method used
in Yang's derivation is too cumbersome, he recalled it as the
longest calculation in his career \cite{20}. Later, less
complicated methods, but which are still too hard to recover, have
been developed to obtain the partition function of the 2D Ising
model \cite{26,27,28,33,34,Syozi}. For works in which other 2D
Ising lattices are considered,  one can give the articles where
average magnetization relations were obtained by Naya for the
Ising model on honeycomb lattice \cite{Naya} and for the
triangular lattice by \cite{Potts}. Of course, there are other
important contributions done on this subject, but it is impossible
to mention all of them in this paper. In other words, it goes well
beyond the scope of the study.

We are now ready to give the general features of the method going
to be used in this paper. Our treatment starts with the recently
obtained relation by Kaya \cite{Kaya}, given as $<\sigma_{0,i}>=$
$ <\tanh[ K(\sigma_{1,i}+\sigma_{2,i}+\dots +\sigma_{z,i})+H]>$.
Here, $K$ is the coupling strength and $z$ is the number of
nearest neighbors. $\sigma_{0,i}$ denotes the central spin at the
$i^{th}$ site while $\sigma_{l,i}$, $l=1,2,\dots,z$, are the
nearest neighbor spins around the central spin. When this relation
is applied to a particular lattice, it produces inevitably odd
spin correlation functions such as the three spins and five spins
correlation functions. At this point it is important to mention
that its applications to the honeycomb
 and the square  lattices are the simplest in that only three spins correlation function appear in the final equation \cite{Kayay},
 making
 the mathematical treatment of the honeycomb and the square lattices are less complicated. The spontaneous magnetization relations have also been
 investigated
 and obtained \cite{Kayay}. In the derivations of the spontaneous magnetization relation of those lattices, we had to use a conjectured mathematical
 form for the corresponding three spin correlation functions, given as $<\sigma_1\sigma_2\sigma_3>=a<\sigma>+(1-a)<\sigma>$. Here $\beta$ denotes the magnetic critical exponent. We think it is also important to point out that the three spin correlation function of the 2D Ising model was considered
 by Baxter \cite{114,Baxter116} for three spins surrounding a triangle. He used the Pfaffian method. Later, a simpler derivation was given by Enting
 \cite{Enting}, who also calculated the three spin correlation for the honeycomb lattice. There are also some other important studies on the subject
 of the three spin correlation functions \cite{104,105,106,Tanaka}  by the cited authors. The important consequence that we get from almost all of them
 is the common physical properties of the three spin correlation function: the three spin correlation function manifestly possesses the same critical
 exponent as the order parameter. As we are going to see in the next section, these physical properties are quite relevant and also apparently
 necessary to describe the three spin correlation function. Therefore, we are going the need to use these properties safely to propose a mathematical functional
 form for the three spin correlation function. In this current work, we will need to make a similar conjecture for the five spin correlation
 functions which are going to be appear in our current investigation. Once again, we are going to use the same physical reasoning as we made for the
 three spin correlation function, to construct the conjectured functional form for the five spin correlation functions.

One might be anxious for the evidence of the power of the approach
which is going to be used in this paper. As a first clue, we
simply want to point out the essence of the method from the
outset. We are always going to start to our treatment with the
relation given for the $<\sigma>$ in the above
paragraph. Exceptional valuable feature of the our approach comes
when this relation is combined with the conjectured odd spins
correlation functions. This combination also provides a more
direct physical insight into making a conjecture about the
mathematical form of the odd spin correlation functions. That is
to say, in the language of our approach, this combination leads to
derive the spontaneous magnetization expression by exploiting the
critical behavior of the order parameter $<\sigma>$. As we are
going to see, our approach turns out to be very useful and
essential in determining the desired analytical  spontaneous
magnetization expression for the long-standing the 3D cubic
lattice problem.

We are going to first consider to calculate an average
magnetization expression for the triangular lattice. For the sake
of confirmation of the validity and relevance of the new
mathematical method, we are going to compare the obtained result
of this paper with the exact spontaneous magnetization relation
for triangular lattice \cite{Codello}. The conventional
mathematical method used to investigate the 2D lattices is so
complicated that, when used to investigate the 3D lattices, any
exact results are still unknown \cite{Zhang,Wu,Rychkov,Guttmann}.
However, the new approach used to investigate to the 2D lattices
with a easily tractable mathematics can also be used in
calculations and investigations of the 3D cubic lattice safely.
Indeed, we are going to deal with the 3D lattices \cite{Perk1,Perk2} here in this paper. It is going to be our main
objective to obtain an analytical spontaneous magnetization
expression for the cubic lattice

This paper is organized as follows. In the next section, the average magnetization of the the triangular lattice is going to be calculated
with an alternative analytic method based on the previously derived average magnetization relation by us \cite{Kayay}.
The obtained expression is also going to be compared by the already available exact result derived by Potts. In the third section,
the average magnetization relation is going to be derived by the same method for the 3D cubic lattice.
The obtained result is going to compared with the available numerical results. In the same section, some discussions
and conclusions are also going to be presented.

\section{The spontaneous magnetization calculation of the triangular lattice}
To investigate and test the validity and relevance of the method which is going to be used in this paper,
we start by considering its application on the triangular Ising lattice. The treatment is going to be very similar to the procedure applied
in the calculation of to the honeycomb and the square lattice cases studied in \cite{Kayay}. To start the calculation,
we need to rewrite the equation introduced in the above introduction once more as,
\begin{equation}
<\sigma_{0,i}>=<\tanh[ K(\sigma_{1,i}+\sigma_{2,i}+\dots +\sigma_{z,i})+H]>.
\end{equation}
For the case of the triangular lattice shown in Fig.1, this equation can expressed quite readily in the absence of external field as,
\begin{figure}[tbph]
    \centering
    \includegraphics[scale=0.50]{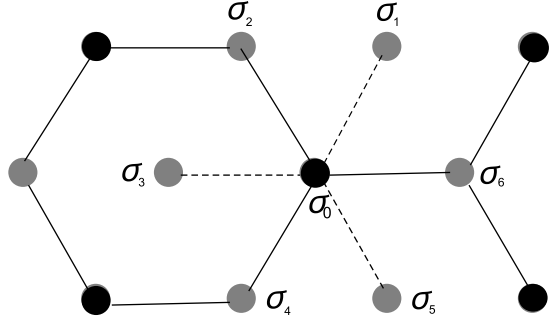}
    \caption{The triangular lattice. The numbers from $1$ to $6$ indicate the spins around the central spin denoted with number zero. The black circles denote the central spins.}
\end{figure}
\begin{equation}
<\!\sigma\!>=
<\tanh[ \kappa(\sigma_1+\sigma_2+\sigma_3+\sigma_{4}+\sigma_{5}+\sigma_{6} )]>,
\end{equation}
where we  keep only numerical indices for notational simplicity. Apparently, this final equation is very similar to the previously used ones
\cite{Kayay} except for extra more spin terms which create eventually some mathematical complication as expected. But, the treatment is going
to be the same. That means, we need to write an equivalent relation which is equal to the hyperbolic tangent function appearing in the last equation.
Thus, recalling the odd functional properties of the function, the following relation can be easily proposed as,
\begin{eqnarray}
&& \tanh[ \kappa(\sigma_{1}+\sigma_{2}+\sigma_{3}+\sigma_{4}+\sigma_{5}+\sigma_{6} )]=A_{t}(\sigma_{1}+\sigma_{2}+\sigma_{3}+\sigma_{4}+\sigma_{5}
+\sigma_{6} )\nonumber\\&&+B_{h}(\sigma_1\sigma_2\sigma_3+ \sigma_1\sigma_2\sigma_4+\sigma_1\sigma_2\sigma_5+\sigma_1\sigma_2\sigma_6
+ \sigma_2\sigma_3\sigma_4+  \sigma_2\sigma_3\sigma_5+ \sigma_1\sigma_3\sigma_6+\nonumber \\&&  \!\!\!\!\!\sigma_3\sigma_4\sigma_5\!+\!  \sigma_3\sigma_4\sigma_6\!+\!
 \sigma_3\sigma_4\sigma_1+ \sigma_4\sigma_5\sigma_6\!+\! \sigma_4\sigma_5\sigma_1+\sigma_4\sigma_5\sigma_2+ \sigma_5\sigma_6\sigma_1\!+\!
  \sigma_5\sigma_6\sigma_2\!\!\!\!\!\!\!\!\!\nonumber \\&&\!\!\!\!\!\!+   \sigma_5\sigma_6\sigma_3+ \sigma_1\sigma_6\sigma_3 + \sigma_1\sigma_6\sigma_4+  \sigma_2\sigma_6\sigma_4+
  \sigma_3\sigma_5\sigma_1 
)+C_{h}(\sigma_1\sigma_2\sigma_3 \sigma_4\sigma_5+\!\!\!\!\!\!\!\!\!\nonumber \\&&\!\!\!\!\sigma_2\sigma_3\sigma_4 \sigma_5\sigma_6\!+\!\sigma_3\sigma_4\sigma_5 \sigma_6\sigma_1+
\sigma_4\sigma_5\sigma_6 \sigma_1\sigma_6\!+\!\sigma_5\sigma_6\sigma_1 \sigma_2\sigma_3\sigma_3\!+\!\sigma_6\sigma_1 \sigma_2\sigma_3\sigma_4).
\end{eqnarray}
Now considering different orientations of the spin variables $\sigma_{i}=\pm 1$, the unknown functions can be obtained after some algebra as,
\begin{eqnarray}
&& A_{t}=\frac{1}{64}\tanh(6K)+\frac{1}{8}\tanh(4K)+\frac{13}{64}\tanh(2K)\nonumber \\&& B_{t}=\frac{1}{32}\tanh(6K)-\frac{3}{32}\tanh(2K)\nonumber
 \\&&C_{t}=\frac{3}{64}\tanh(6K)-\frac{1}{8}\tanh(4K)+\frac{7}{64}\tanh(2K).
\end{eqnarray}
If Eq. (3) is substituted into Eq. (2), after taking into account the equivalent odd spins correlations, it leads to,
\begin{eqnarray}
&&<\!\sigma\!>=6A_{t}<\!\sigma\!>+B_{t}[5<\sigma_1\sigma_2\sigma_3>+13<\sigma_1\sigma_{3}\sigma_4>\nonumber \\&&+2<\sigma_1\sigma_3\sigma_5>]+
6C_{h}<\sigma_1\sigma_2\sigma_3\sigma_4\sigma_5>.
\end{eqnarray}
Apparently, for three spins correlations appearing in the last equation, the same conjectured functional forms used in the reference \cite{Kayay}
can be used safely as they lead to obtain almost exact spontaneous magnetization relations for the honeycomb and the square lattices. Furthermore, considering the 
last equation, it is apparent that the five spins correlation functions must have also the same critical behavior as the order parameter due to the order parameter vanishes for the values of $K<K_{h,c}$. This final property of the five spins correlation functions indicates that it can also be written as a function of
the order parameter $<\sigma>$ . Of course, as in the case of three spins correlation functions, it is almost impossible to obtain exact functional expression for 
 either three spins or five spins correlation in terms of the order parameter. But, it is always possible to conjecture up a functional form for them. 
Since the construction of the odd spins correlation functions is the one of the important parts of this paper, they deserve more rigorous mathematical explanation. Unfortunately, it has been proven that it is not easy, if not impossible \cite{Gut,Zhang}. However, due to structure of the last equation, it is straight forward 
to claim that the odd spins correlation functions can be a function of the order parameter $<\sigma>$. Furthermore, from physical point of view, we may also claim 
that they all must obey the same critical behavior of the order parameter. Taking into account these facts and for the sake of simplicity,  
we may assume that the same conjectured form for three spins correlation function used in cite{Kayay} might be also the proper choice as well for the five spin 
correlation function. Under these considerations, we can express the three and five spin correlation functions as,
\begin{eqnarray}
\!\!\!\!\!\!\!\!&&<\sigma_1\sigma_2\sigma_3>=a_{t,1}<\sigma>+(1-a_{t,1})<\sigma>^{1+\beta^{-1}}\nonumber \\&&<\sigma_1\sigma_3\sigma_4>=
a_{t,2}<\sigma>+(1-a_{t,2})<\sigma>^{1+\beta^{-1}}\nonumber \\&&<\sigma_1\sigma_3\sigma_5>=a_{t,3}<\sigma>+(1-a_{t,3})<\sigma>^{1+\beta^{-1}}\nonumber
 \\&&<\sigma_1\sigma_2\sigma_3\sigma_4\sigma_5>=a_{t,4}<\sigma>+(1-a_{t,4})<\sigma>^{1+\beta^{-1}}.
\end{eqnarray}
Substituting the last relations into Eq. (5), it can be written as,
 \begin{eqnarray}
&&<\sigma>=[6A_{t}+(5a_{h,1}+13a_{t,2}+2a_{h,3})B_+6C_{t}a_{h,4}]<\sigma>+\nonumber \\&&[B_{t}( 5(1-a_{t,1}){h}\!+\!13(1-a_{t,2})\!+\!2(1-a_{t,3}))\!+\!6C_{t}(1-a_{t,4})]<\!\sigma\!>^{1+\beta^{-1}}.
\end{eqnarray}
In the last equation, there are four unknown parameters that need to be determined properly. To this end, it is relevant to consider the restrictions which
these parameters need to satisfy. First, all of these parameters have to assume positive real values less than unity. Second, considering the distances
between spins, it is easy to see that $a_{t,1}>a_{t,2}>a_{t,3}$  and $a_{t,4}$ must have the values less that the values of all the other three parameter because 
it is used in expressing five spins correlation function. From the obtained values for the parameters in the \cite{Kayay}, one can intuitively assume that
all of the four parameters take values close to each other. Under these considerations, we introduce the following statements to present the last
equation with a single parameter equation. Apparently, we need to define the new parameter in terms of the four parameters. Thus,
if we name $z_{1}=5a_{t,1}+13a_{t,2}+2a_{t,3}$ and $z_{2}=6a_{t,4}$, one can easily see that $z_{2}\simeq \frac{6}{20}z_{1}$ due to the assumption
that the values of all of the four parameters are approximately equal to each other. Under this consideration the last equation can be expressed as,
\begin{eqnarray}
&&<\sigma>=[6A_{h}+(B_{h}+\frac{3}{10}C_{h})z_1]<\sigma>+\nonumber \\&&[20B_{h}+6C_{h}-(B_{h}+\frac{3}{10})z_{1}]<\sigma>^{1+\beta^{-1}}.
\end{eqnarray}
From this equation, one can easily obtain in a self-consistent manner, the value of $z_{1}$ at critical point as,
\begin{equation}
1-[6A_{h}+(B_{h}+\frac{3}{10}C_{h})z_1] |_{K=K_{c}}=0.
\end{equation}
Substituting the critical coupling strength  value for triangular lattice $K_{c}=0.2747$ into the last relation, the value of $z_{1}$ is calculated
as $z_{1}=16.13$.
Finally, from Eq. (7), one can easily obtain the following expression for the spontaneous magnetization of the triangular lattice as,
\begin{equation}
<\sigma>=\left[\frac{1-[6A_{h}+(B_{h}+\frac{3}{10}C_{h})z_1]}{20B_{h}+6C_{h}-(B_{h}+\frac{3}{10})z_{1}}\right]^{\beta}.
\end{equation}
In Fig. 2, the spontaneous magnetization expression given in Eq. (10) and the exact result \cite{Potts,Codello},
given as
\begin{equation}
<\sigma>=\frac{(3-e^{4K})(1+e^{4K})^{3}}{(3+e^{4K})(1-e^{4K})^{3}}
\end{equation}
\begin{figure}[tbph]
    \centering
    \includegraphics[scale=0.55]{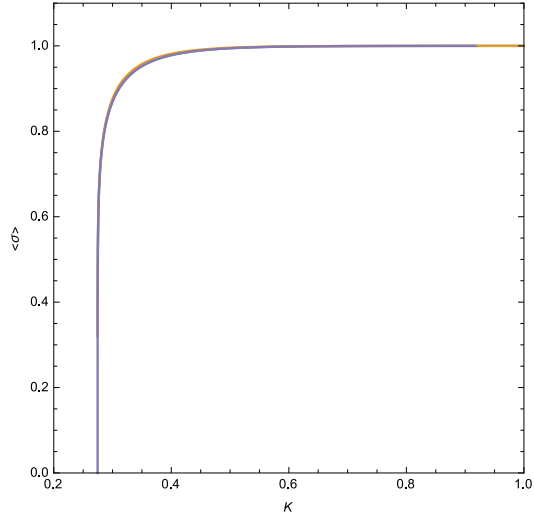}
    \caption{The comparison of the obtain spontaneous magnetization in this paper with the exact result. It is important to notice that they are
    almost equivalent.}
\end{figure}
are plotted. As seen from the figure, they are almost equivalent. This  means that an almost exact result is obtained by avoiding the very daunting
and cumbersome conventional method used previously to calculate the 2D lattice spontaneous magnetization. In addition, the relevance of the
conjectured odd spins correlation functions are testified and confirmed by this excellent agreement. Hence, the conjectured functional form 
can be used safely regardless of the lattice types for the 2D lattice Ising 
model. At this point, it is important to mention once more that the mathematical and physical approaches used in this paper are a lot more easier 
than the calculation with the conventional calculation methods. We think, therefore, that it may help us in the calculation of the spontaneous magnetization for the 3D lattices. If we recall that all attempts have been unsuccessful since now
for the analytical treatment of the 3D lattices. We are now in a position to deal with the long-standing unsolved problem.
Due to the relevance and tractability
of the method used in this paper, we are really hopeful that one can obtain the analytic expression for the spontaneous magnetization for the 3D lattices 
with almost in the same manner as used for the 2D lattices. To this end, we are going to consider the cubic lattice Ising model in the following section.

\section{The spontaneous magnetization calculation for the cubic lattice}
In this section, we are going to derive an analytical expression for the spontaneous magnetization of the 3D cubic lattice which is unknown up to now,
since its derivation is not possible with the conventional method used for the 2D lattices. We now know, however, that the approach used in this
paper is quite relevant and accurate for the calculation of the spontaneous magnetization relation for all of the 2D lattices. Therefore,the approach
used in this paper and \cite{Kayay} can be applied equally well to the cubic lattice. For the spirit of present approach, it is natural to start with
the equation Eq. (1)  in the absence of the external field, which can be expressed for the cubic lattice (see Fig. 3) as
\begin{equation}
<\sigma>=
<\tanh[ \kappa(\sigma_1+\sigma_2+\sigma_3+\sigma_{4}+\sigma_{5}+\sigma_{6} )]> .
\end{equation}
As used above, our next step is to write an equivalent function for the hyperbolic tangent appearing in this equation to relate the spontaneous
magnetization with the other correlation functions. To this end, we find that the following relation is convenient to satisfy the desired equivalence,
\begin{eqnarray}
&&\!\!\!\!\!\!\!\!\!\!\!\!\!\!  \tanh[ \kappa(\sigma_{1}\!+\!\sigma_{2}\!+\!\sigma_{3}\!+\!\sigma_{4}\!+\!\sigma_{5}\!+\!\sigma_{6})]=A_{c}[\sigma_{1}\!+\!\sigma_{2}\!+\sigma_{3}\!+\!\sigma_{4}\!+\!\sigma_{5}+
\sigma_{6} ]\!+\!\nonumber \\&&\!\!\!\!\!\!\!\!\!\!\!\!\!\! B_{c}[\sigma_{1}\sigma_{2}\sigma_{3}+ \sigma_{2}
\sigma_{3}\sigma_{4}+\sigma_{4}\sigma_{1}\sigma_{2}+ \sigma_{1}\sigma_{3}\sigma_{4}+ \sigma_{2}\sigma_{3}\sigma_{5}+
\sigma_{2}\sigma_{3}\sigma_{6}+ \sigma_{3}\sigma_{4}\sigma_{5}+\nonumber \\&&\!\!\!\!\!\!\!\!\!\!\!\!\!\! \sigma_{3}\sigma_{4}\sigma_{6}+\sigma_{2}\sigma_{4}\sigma_{5} +
 \sigma_{1}\sigma_{4}\sigma_{6}+
\sigma_{1}\sigma_{2}\sigma_{5}+\sigma_{1}\sigma_{2}\sigma_{6}+\sigma_{1}\sigma_{3}\sigma_{6}+ \sigma_{2}\sigma_{4}\sigma_{5} +\nonumber\\&&
\!\!\!\!\!\!\!\!\!\!\!\!\!\! \sigma_{2}\sigma_{4}\sigma_{6}+  \sigma_{5}\sigma_{6}\sigma_{1}+  \sigma_{5}\sigma_{6}\sigma_{2}+ \sigma_{5}\sigma_{6}\sigma_{3}+
 \sigma_{5}\sigma_{6}\sigma_{4} ]+C_{c}[\sigma_{1}\sigma{2}\sigma_{3}\sigma_{4}\sigma_{5}+ \nonumber\\&&\!\!\!\!\!\!\!\!\!\!\!\!\! \! \sigma_{1}\sigma{2}\sigma_{3}\sigma_{4}\sigma_{5}+
 \sigma_{5}\sigma_{6}\sigma_{1}\sigma_{2}\sigma_{3}+\sigma_{5}\sigma_{6}\sigma_{2}\sigma_{3}\sigma_{4}+
 \sigma_{5}\sigma_{6}\sigma_{1}\sigma_{2}\sigma_{4}+ \sigma_{5}\sigma_{6}\sigma_{1}\sigma_{3}\sigma_{4} ].
\end{eqnarray}
\begin{figure}[tbph]
    \centering
    \includegraphics[scale=0.50]{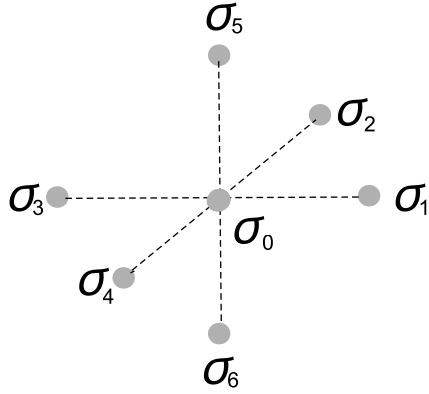}
    \caption{The six surrounding spins for the cubic lattice around the central spin denoted by zero. $\sigma_{1}, \sigma_{2}, \sigma_{3}$ and $\sigma_{4}$ are on $xy$ plane while $\sigma_{5}$ and $\sigma_{6}$ are on the $z$ axis.}
\end{figure}
Now considering different orientations of the spin variables $\sigma_{i}=\pm 1$, the unknown functions can be obtained after some algebra as,
\begin{eqnarray}
&& A_{t}=\frac{1}{64}\tanh(6K)+\frac{1}{8}\tanh(4K)+\frac{13}{64}\tanh(2K)\nonumber \\&& B_{t}=\frac{1}{32}\tanh(6K)-\frac{3}{32}\tanh(2K)\nonumber
 \\&&C_{t}=\frac{3}{64}\tanh(6K)-\frac{1}{8}\tanh(4K)+\frac{7}{64}\tanh(2K).
\end{eqnarray}
Now, taking into account the equivalent spin orientations which apparently produce the same correlation functions, the following equation can be
easily written from the Eq. (12) as,
\begin{eqnarray}
&&\!<\!\sigma\!>=6A_{c}<\!\sigma\!>+B_{c}[4<\!\sigma_1\sigma_2\sigma_3\!>\!+\!8<\!\sigma_1\sigma_{2}\sigma_6\!>\!\!\!+4<\!\sigma_1\sigma_3\sigma_6\!>\nonumber \\&&+4<\sigma_1\sigma_5\sigma_6>]\!+C_{t}[2<\!\!\sigma_1\sigma_2\sigma_3\sigma_4\sigma_5\!>\!+
\!4<\!\sigma_5\sigma_6\sigma_1\sigma_2\sigma_3\!>].
\end{eqnarray}
From this equation it is clear that the critical behavior of the three spins correlation functions and the five spins correlation functions must have the same critical feature as the order parameter. Hence, for all practical purposes the conjectured form for the correlation functions for
the 2D lattices can be also have the same functional form for the cubic lattice. This means that  the following relations for those correlation functions
appearing in the last equation can be relevant and safe functional forms. Thus,
\begin{eqnarray}
&&<\sigma_1\sigma_2\sigma_3>=a_{c,1}<\sigma>+(1-a_{c,1})<\sigma>^{1+\beta^{-1}}\nonumber \\&&<\sigma_1\sigma_2\sigma_6>=a_{c,2}<\sigma>+
(1-a_{c,2})<\sigma>^{1+\beta^{-1}}
\nonumber \\&&<\sigma_1\sigma_2\sigma_3>=a_{c,3}<\sigma>+(1-a_{c,3})<\sigma>^{1+\beta^{-1}}\nonumber
\nonumber\\&&<\sigma_1\sigma_5\sigma_6>=a_{c,4}<\sigma>+(1-a_{c,4})<\sigma>^{1+\beta^{-1}}
\nonumber \\&&<\!\sigma_1\sigma_2\sigma_3\sigma_4\sigma_6\!>=a_{c,5}<\!\sigma\!>+(1-a_{c,5})<\!\sigma\!>^{1+\beta^{-1}}
\nonumber \\&&<\!\sigma_1\sigma_2\sigma_3\sigma_5\sigma_6\!>=a_{c,6}<\!\sigma\!>+(1-a_{c,6})<\!\sigma\!>^{1\!+\!\beta^{-1}}.
\end{eqnarray}
Substituting these functions into Eq. (15)  leads to
\begin{eqnarray}
&&<\!\sigma\!>=[6A_{c}\!+\![B_{c}(4a_{c,1}\!+\!8a{c,2}\!+\!4a_{c,3}\!+\!4a_{c,4})\!+\!C_{c}(2a_{c,5}\!+\!4a_{c,6})]\!<\!\sigma\!>\nonumber \\&&+[B_{c}(4(1-a_{c,1})+8(1-a_{c,2})+4(1-a_{c,3})+4(1-a_{c,4}))\nonumber \\&&+C_{c}((1-a_{c,5})4(1-a_{c,6}))]<\sigma>^{1+\beta^{-1}}.
\end{eqnarray}
We now find it convenient to define  $z_{1}=4a_{c,1}+8a_{c,2}+4a_{c,3}+4a_{c,4}$ and $z_{2}=2a_{c,5}+4a_{c,6}$. If we assume the values of $a_{c,i}$
are close to each others as assumed for the triangular lattice case, $z_{2}$ can be approximated as $z_{2}\simeq\frac{6}{20}z_{1}$.
Taking into acount these final remarks, the last equation may be written as,
\begin{eqnarray}
&&<\sigma>=[6A_{c}+(B_{c}+\frac{3}{10}C_{c})z_{1}]<\sigma>\nonumber\\&&+[20B_{c}+6C_{c}-(B_{c}+\frac{3}{10}C_{c})z_{1}]<\sigma>^{1+\beta^{-1}}.
\end{eqnarray}
From this equation the value of the $z_1$ can be obtained self-consistently, by considering the criticality of the spontaneous magnetization at the
critical point $K_{c}=0.2216$ as
\begin{equation}
1-[6A_{c}+(B_{c}+\frac{3}{10}C_{c})z_{1}]|_{K=K_{c}}=0.
\end{equation}
\begin{figure}[tbph]
    \centering
    \includegraphics[scale=0.55]{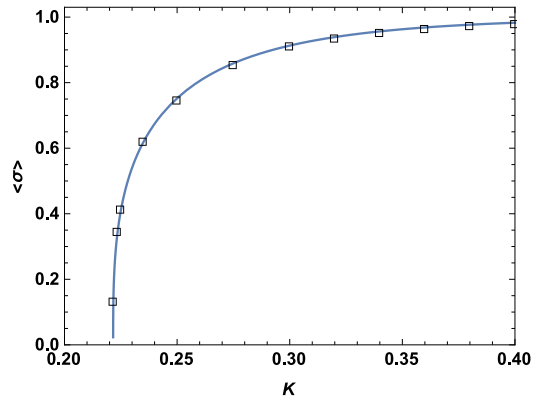}
    \caption{The plot of the spontaneous magnetization for the cubic lattice given by Eq. (16). The symbol $\square$ indicates the numerical data
    calculated in Tapalov paper. }
\end{figure}
This equation produces $z_{1}=9.6105$. Now, it is straightforward to obtain the spontaneous magnetization relation for the cubic lattice from the
Eq. (18) as,
\begin{equation}
<\sigma>=\left[\frac{1-[6A_{c}+(B_{c}+\frac{3}{10}C_{c})z_{1}]}{20B_{c}+6C_{c}-(B_{c}+\frac{3}{10}C_{c})z_{1}}\right]^{\beta}.
\end{equation}
It is now quite natural to ask about the relevance of the obtained spontaneous magnetization relation for the cubic lattice. There is, of course, no immediate answer for this question since there is no previously obtained analytic
expression for the spontaneous magnetization for the cubic lattice. On the other hand, there are some numerical calculation results for the 3D cubic lattice \cite{Tapa}. So it is convenient to plot the graphics of the Eq. (20) together with the numerical data. Indeed, Fig. 4 is plotted to compare Eq. (20) with the numeric data obtained by Tapalov and Bl\"{o}te \cite{Tapa}. As seen from the
figure, the numeric data points are totally in agreement with the curve of Eq. (20). Furthermore, in the paper of Tapalov, an empirical or numerical
relation is introduced which is valid in the interval $0.0005<t<0.26$. The relation expressed by the  Eq. (10) of Tapalov paper is given as,
 \begin{equation}
<\sigma>=t^{0.32694}[1.691904-0.343577t^{0.5084}-0.42572t]
\end{equation}
where $t=1-\frac{K_{c}}{K}$, $K_{c}=0.2216544$ was used by Tapalov. The value of the critical exponent for $\beta$ is obtained as $0.32694$ in the same paper. These critical values are also obtained with very high accuracy in \cite{Ferrenberg,Hu}. At this point it is also important to notice that Tapalov's numerical relation is confirmed by the research paper \cite{Xie}, in which a new coarse grain tensor renormalization group method based on the higher order singular value decomposition method is used.
Therefore, we think that the Eq. (10) of Tapalov paper is relevant enough to safely compare the spontaneous magnetization relation obtained in this paper.
\begin{figure}[tbph]
    \centering
    \includegraphics[scale=0.55]{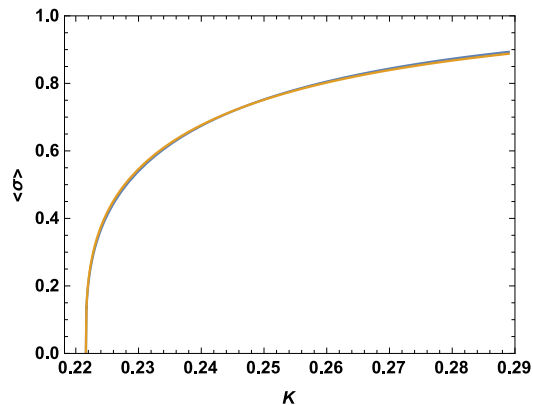}
    \caption{The comparison of Eq. (16) with the Tapalov's Eq. (10).}
\end{figure}
Fig. 5 is plotted for comparing  Eq. (20) with the Eq. (10) of Tapalov. As seen from the figure, these two relations are almost equivalent within the claimed range of the Eq. (10) of the Tapalov's paper. We, therefore, think that the spontaneous magnetization relation expressed in Eq. (20) of this paper is quite relevant result to be safely used wherever it is necessary. In other words, we now have a relevant  and almost exact analytical expression for the 3D cubic lattice. Therefore, Eq. (20) can be considered as a very important development for the analytic treatment of the 3D Ising model. 
In summary, we have achieved to derive an relevant and almost exact formula for the spontaneous magnetization of the cubic lattice almost 70 years 
after the the square lattice solution of Yang. In addition, from the obtained result of this paper, we can also claim that the conjectured functional 
forms for the odd spins correlation functions around the central spin can possibly considered as the almost exact expression for those correlation functions. This means that one can safely use these conjectured forms wherever they are needed.
\section{Conclusion and Discussion}
In this work we have derived the spontaneous magnetization relations for the triangular and for the cubic lattice Ising model by considering a novel
mathematical approach. The new approach is based on the previously obtained general equation, which relates the spontaneous magnetization with the
corresponding odd spin correlation functions \cite{Kaya}. Taking into account the criticality properties  of the spontaneous magnetization, we are forced to assume that the odd spins correlation functions have to also obey the same criticality of the spontaneous magnetization. Exploiting these similarities,
the mathematical functional forms for the odd spins correlation functions are conjectured. Using these conjectured functions, the spontaneous magnetization relations for those lattices are obtained. Comparing the obtained spontaneous magnetization expressions with the already available results, the physical relevance of the obtain magnetization expressions are tested. Since the agreement between our results and the previously obtained results are almost the same, we may claim that the method used in this paper is quite relevant and is a safe method to calculate the spontaneous magnetization of the Ising lattices in general. In other words, the approach used in this paper simplifies not only the mathematical work needed for the exact treatments, but also it retains almost exactly the same essential critical features of the problem. It is also important
to mention once more that there has been no analytic spontaneous magnetization relation for the cubic lattice in 3D up to now. The analytic spontaneous magnetization expressed by Eq. (16) of this paper is, therefore, very important and valuable. In addition, we can also claim from the result of this paper that the
 conjectured mathematical form for the odd spins correlation functions can be considered as excellent approximations (or almost exact expressions).
 Furthermore, because of the simplicity and generality of the method, it can be also used the calculation of the spontaneous magnetization of the other 3D lattices. Maybe even under some restrictions, it can be used to calculate the spontaneous magnetization in the presence of the external field.

\end{document}